# Quantum probability updating from zero prior (by-passing Cromwell's rule)

Irina Basieva[a,b*], Emmanuel Pothos[b], Jennifer Trueblood[c], Andrei Khrennikov[a], and Jerome Busemeyer[d]

[a] Linnaeus University, Universitetplatsen 1 351 95 Växjö, Sweden
[b] Department of Psychology, City University London, UK
[c] Department of Psychology, Vanderbilt University, USA
[d] Department of Psychology, Indiana University, USA

**Abstract**

Cromwell's rule (also known as the zero priors paradox) refers to the constraint of classical probability theory that if one assigns a prior probability of 0 or 1 to a hypothesis, then the posterior has to be 0 or 1 as well (this is a straightforward implication of how Bayes's rule works). Relatedly, hypotheses with a very low prior cannot be updated to have a very high posterior without a tremendous amount of new evidence to support them (or to make other possibilities highly improbable). Cromwell's rule appears at odds with our intuition of how humans update probabilities. In this work, we report two simple decision making experiments, which seem to be inconsistent with Cromwell's rule. Quantum probability theory, the rules for how to assign probabilities from the mathematical formalism of quantum mechanics, provides an alternative framework for probabilistic inference. An advantage of quantum probability theory is that it is not subject to Cromwell's rule and it can accommodate changes from zero or very small priors to significant posteriors. We outline a model of decision making, based on quantum theory, which can accommodate the changes from priors to posteriors, observed in our experiments.

## 1. Introduction

Luce is known for many seminal contributions in psychology, yet he still considered the modelling of uncertainty and vagueness as one of the "unresolved conceptual problems of mathematical psychology" (the text is part of the title of Luce, 1997). This work considers a fundamental aspect of probabilistic inference, probability updating.

According to classical probability (CP), our belief in different hypotheses should be updated using Bayes's law, which states that

$$p(H_i | D) = \frac{p(D|H_i) p(H_i)}{\sum_j p(D|H_j) p(H_j)}. \qquad (1)$$

Here $H_i$ is a particular hypothesis and $D$ is the observed data. In the set-theoretical paradigm, Eq. (1) is equivalent to

$$p(H_i | D) = \frac{p(H_i \cap D)}{\sum_j p(H_j \cap D)} \qquad (1')$$

The term $p(H_i \cap D)$ is the joint probability of a hypothesis and the data. As can be seen by the above formula, when holding the data fixed, Bayes rule is just a special case of Luce's choice rule, where $p(H_j \cap D)$ serves as the strength of an alternative. When viewed this way, Bayes rule must satisfy the choice axiom as well.

There is little doubt that Bayes's law is successful on many occasions, as is evident by the many CP theory psychological models providing excellent fits (Griffiths et al., 2010; Oaksford & Chater, 2007; Tenenbaum et al., 2011) and the widespread usage (e.g., in machine learning, management, economics) of causal reasoning systems based on so-called Bayes Nets.

Bayes rule must satisfy a very important property: when the prior is 0, then regardless of the data that we observe, the posterior probability, $p(H_i | D)$ as calculated by Bayes's law, has to be 0 as well.

---

[*] Corresponding author (irina.basieva@gmail.com)

Likewise, when the prior is 1, the posterior has to be 1 as well. These observations have been called Cromwell's rule, honoring Oliver Cromwell and his words to the members of the synod of the Church of Scotland "to think it possible that you may be mistaken" (Carlyle, 1885). The argument goes that (e.g., if one cares to stay open-minded) one should assign some small possibility to even the most improbable state of affairs. Otherwise, however much evidence is subsequently accumulated in favor of the zero-prior possibility, one will be trapped with all-zero posteriors – we call this the zero priors trap. The trap has roots in the stringent linearity of Bayes's law; updated probabilities are linearly dependent on the priorsThe main objective of this work is to suggest that the requirement from Bayes's law may be too stringent, at least in some occasions.

In human decision making, how can the impossible become possible? In this article, we consider belief updating (in the context of decision making). We discuss a simple experimental paradigm, which involves the presentation of options for a hypothesis/ problem, some of which are initially impossible/ extremely unlikely. We then create a situation, by providing additional information, which makes some of these initially impossible options likely. Do participants actually update their beliefs in a way that is inconsistent with Bayes's law? That is, do participants produce evaluations of posterior probabilities that exceed the prior probabilities by amounts that go beyond what is allowed by Bayes's law? An interesting angle to this question concerns creativity, one of the most intriguing features of human cognition. We do not mean to define or evaluate creativity in this paper. However, whatever creativity is about, it must also involve a component of eventually recognizing as likely possibilities that had zero or very low probabilities, to start with. That is, we suggest that, to be "creative", a person has to transcend Cromwell's rule, in order to assess options, which had been dismissed or not considered before.

We show that there are situations when human decision making is not consistent with Cromwell's rule. Then we argue that the zero-priors trap can be naturally avoided if one employs quantum probability theory (QPT), in particular, the QPT rule for updating probabilities, the von Neumann – Lüders' rule. QPT allows jumps in belief, that is, discontinuous changes in belief states (resulting from measurements).

We borrow from QPT, initially developed for quantum mechanics, the rules for how to assign probabilities to events, without any of the physics. QPT is applicable in any situation where there is a need to formalize uncertainty and, more recently, it has been used to construct cognitive models, especially in decision making (Pothos & Busemeyer, 2013; Busemeyer & Bruza, 2012; Busemeyer et. al., 2006; Busemeyer et.al., 2011; Khrennikov, 2003; Haven & Khrennikov, 2012). Such models have enjoyed good descriptive success, especially for findings which have been paradoxical from a CP perspective. This is due to the fact that many of such so-called paradoxes, such as order effects (Trueblood & Busemeyer, 2011), the disjunction effect (Hofstader, 1983; Croson, 1999), and violations of the law of total probability (Conte et.al., 2007) are conveniently resolved in QPT. It provides an entirely different (compared to the standard CP one) paradigm for probabilistic inference, where events and probabilities are represented by vectors in Hilbert space and Hermitian operators. We shall outline the instruments useful to overcome the zero-priors paradox in the next section. This can also account for violations of the law of total probability, which has been impressively demonstrated in a number of experiments (Shafir & Tversky,1992; Tversky & Shafir, 1992; Tversky & Kahneman, 1983; Busemeyer et.al., 2009). Our goal is to experimentally test the validity of Bayesian updating of zero or extremely small prior and to apply QPT to the results.

A natural domain to look for a suitable decision making paradigm to test Cromwell's rule is detective stories, where the criminal turns out to be someone completely unexpected (as in Dostoyevsky's *The Brothers Karamazov,* where, in full compliance with CP rules, prior beliefs about the actual criminal were so weak, that his complete confession together with other hard evidence failed to convince the court) or not even a person (as in Poe's *The Murder in the Rue Morgue).* In the latter case, the available prior options do not exhaust the probability space. As briefly noted, a situation when a previously non-existent option suddenly comes into consideration seems related to creativity. But this is precisely the problem: some possibilities cannot be, in principle, listed from the very beginning, since some of them are totally novel and have not been encountered before. For example, it may happen that the moon contains some minerals that are not present on earth. We are not able to assign even a tiny probability to such a possibility, in advance.[1] Here the classical Bayesian analysis seems simply inapplicable.

The standard way to deal with such situations in classical probability theory is additive smoothing, but the problem with it (from the viewpoint of our challenge) is that it is applied a posteriori. For example, after we discover a new mineral, we reshuffle the prior probabilities adding a "pseudo-count" to all the priors, so that no option on the (a posteriori!) list has a zero prior. There are other approaches (Hoppe, 1984), such as the Pólya urn model, directly enforcing increase in probability for a less probable event each time the more probable event happens. For example, starting with the urn containing only (or mostly) black balls, we add a white ball each time a black ball is drawn. A variation most interesting to us considers adding a ball of some new color (instead of white) each time a black ball is drawn. This is analogous to adding a new option to the list. Still, the question of "creating" the new colors or new options remains open. The bottom line is that creativity is at odds with the linear dependence of the posterior on the prior. How fundamental is this feature of Bayesian updating in human cognition?

The validity of Bayesian updating has been questioned before, for example, by Van Wallendael and Hastie (Robinson & Hastie, 1985; Van Wallendael, 1989; Van Wallendael & Hastie, 1990) who noted that, upon receiving information about one hypothesis, people tend to revise only the corresponding probability and leave their other estimates untouched (so that the total fails to equal one). Among alternative approaches (in CP), aiming to describe this anomaly, there is Shafer's representation of belief states (Shafer, 1976) where possible options are grouped (so that one has beliefs about the groups rather than individual hypotheses). In Section 4 we briefly consider whether this approach is useful for solving the zero priors paradox. Fundamentally, Shafer's theory of belief functions is consistent with CP, which makes its application to the present problem limited. In contrast, our main proposal is that QPT is an appropriate way to generalize CP (and, in particular, allow for nontrivial updating of zero, or extremely small, priors). Also, approaches beyond CP theory are not necessarily all quantum, and there are ones that are even more non-classical, compared to QPT. For example, there are probability models that account for "negative probabilities" regularly encountered both in physics and cognitive psychology (Acacio de Barros, 2014; Acacio de Barros & Oas, 2014).

In the present paper, we restrict ourselves to a simple experimental setup, where all suspects are listed beforehand. This allows us a more direct approach and provides for easier analysis. We demonstrate that the experimental probabilities given by participants in the course of a crime-solving test strongly violate Bayesian updating. In fact, in 20% of cases participants went from zero prior to high confidence, in a single step of updating. Such results can be described by QPT.

---

[1] Even if we reserve some portion of our belief to be responsible for all highly improbable possibilities, this does not help us in creating a new option out of nothing, when we have a set of (not explicitly defined) improbable possibilities rather than a rigid set of explicit options.

## 2. Principles of Quantum-like Updating

The quantum approach to probability updating involves the formalism of a complex Hilbert space *H* (for representing belief states) and the theory of Hermitian operators (for representing observables, including decision operators).

In QPT, (pure) belief states are represented by normalized vectors in a Hilbert space *H* (a complete vector space, where the scalar product between vectors is defined and denoted here as angle brackets). A scalar product determines the norm of any vector $\psi$ on *H* as $\|\psi\|=<\psi,\psi>$. Vector components are, in general, complex numbers. Given a complex number $z=u+iv$, we denote its conjugate as $z^*$, $z^*=u-iv$. The number of vector components is, in general, infinite. Here we confine ourselves to an *m*-dimensional Hilbert space, where an orthonormal basis $\{e_1, e_2, e_m\}$ is chosen. *H* can then be represented as a Cartesian product $C^m$, where the scalar product $<\psi,\phi>$ of vectors $\psi=c_1e_1+c_2e_2+\ldots c_me_m$ and $\phi=k_1e_1+k_2e_2+\ldots k_me_m$ is then given by $c_1k_1^*+c_2k_2^*+\ldots c_mk_m^*$. Operators are simply Hermitian matrices. A square matrix *A* with elements $a_{ij}$ is Hermitian if it is equal to its conjugate transpose, that is $a_{ij}=a_{ji}^*$.

Eigenvalues of Hermitian operator are the various possible values that can be obtained with a measurement. Immediately after a measurement, the state of the system is given by a projection of the initial state to the eigen-subspace corresponding to the eigenvalue that was obtained, as the result of the measurement. The latter statement is known as the von-Neumann-Lüders projection postulate.

When observables do not commute, i.e., incompatible observables, the features of quantum probability updating differ crucially from the well-known features of Bayesian probability updating. Instead, the von-Neumann-Lüders postulate is used in QPT to update the prior state $\psi_0$ by the means of projector operator $E_\theta$ to a new state:

$$E_\theta\psi_0 / \|E_\theta\psi_0\|. \qquad (2)$$

In classical probability theory there is the set of hypotheses or "states" $\Theta=\{\theta_1, \theta_2, \ldots \theta_m\}$, which are mutually exclusive (only one of them is actually true) and exhaust the whole probability space, so that the sum of all probabilities $\pi(\theta_i)$ is one. More generally, when $\Theta$ is not necessarily discrete, one can speak about a probability density $\pi(\theta)$ on $\Theta$ rather than discrete probabilities $\pi(\theta_i)$.

Constructing a big enough state space is an important first step. From the start, we have to account even for the most inconceivable possibilities by introducing the corresponding states into consideration (e.g., that the moon is made of green cheese). In contrast, in the quantum model, see below, we are free to assign to them zero priors.

For a random variable *X* taking values from the set $\{x_1, x_2, \ldots x_n\}$, one can specify the probability distribution $p(x|\theta)$ for each state $\theta$. We use two different letters $\pi$ and $p$ to distinguish between probability distributions for hypotheses $\Theta$ and information *X*. Now, if the random variable *X* is measured, one can update the prior probability distribution on the basis of information gained from this concrete result of measurement, say $x_i$. The classical probability updating gives us $\pi(\theta|x)$ according to the Bayes rule (leaving probabilities 0 and 1 invariant), as:

$$\pi(\theta|x_i)= p(x_i|\theta)\, \pi(\theta)/ p(x_i). \qquad (3)$$

Consider two observables $\Theta$ with values $\theta_1, \theta_2, \ldots \theta_m$ and *X* with values $x_1, x_2, \ldots x_n$. The first one corresponds to the initial set of hypotheses and the second one to some additional information (which will be used for probability updating). In QPT, the observables are represented by Hermitian operators, which we can denote as $\Theta$ and *X* as well, with eigenvalues $\theta$ and *x*, correspondingly. Then,

$$\Theta = \sum \theta E_\theta \text{ and } X = \sum x F_x, \qquad (4)$$

where summation is over all possible values $\theta$ or $x$, and $(E_\theta)$ and $(F_x)$ are orthogonal projectors corresponding to the eigen-subspaces of these operators.

For a particular person, the initial mental representation relevant to a situation is given by a *belief state* $\psi_0 \in H$. Within this state, all subjective probabilities for different states of nature are already encoded. They can be extracted by performing direct measurements of $\Theta$ and taking the square of the norm of the resulting vector:

$$\pi(\theta) = \langle E_\theta \psi_0 | \psi_0 \rangle = \|E_\theta \psi_0\|^2 \qquad (5)$$

This observation procedure models decision making about prior possibilities of $\Theta$ contained in $\psi_0$.[2]

We want to update the probabilities of $\theta$ on the basis of additional information from measurement of the *X*. By using Lüders rule (2), which is the quantum rule for conditional probabilities, we get

$$\pi(\theta | x) = \frac{\langle E_\theta F_x \psi_0 | F_x \psi_0 \rangle}{\|F_x \psi_0\|^2}. \qquad (6)$$

The result is not constrained to coincide with Bayesian probability updating. We shall address the question of when quantum probability updating coincides with classical updating elsewhere.

Let us explore how these ideas apply to a situation, analogous to the one in our empirical investigation. Suppose that some crime case is under investigation by a police officer. He has the list of suspects $\{\theta_1, \theta_2, \ldots \theta_m\}$. A "classically thinking commissar" assigns some probabilities that the crime was done by these suspects, $\pi(\theta_i)$, $i = 1, 2, \ldots m$. Then, in the process of investigation he obtains new pieces of evidence related to this crime, which are encoded by some number *x*. For each suspect *i* he assigns probability $p(x|\theta_i)$ that these pieces of evidence given by *x* can correspond to the case that the crime was done by suspect $\theta_i$. Finally, he applies Bayes' formula (3) to get the probability $\pi(\theta_i|x)$ that $\theta_i$ is really responsible for this crime. If some person $\theta$ was not present in the initial list $\{\theta_1, \theta_2, \ldots \theta_m\}$, i.e., $\pi(\theta)=0$, the commissar would never get a nontrivial $\pi(\theta_i|x)$.

A "quantum thinking commissar" prepares the initial belief state $\psi_0$. It can be represented as superposition of belief states corresponding to the suspects in the initial list:

$$\psi_0 = \sum_i c_i e_{\theta_i}, \quad \sum_i |c_i|^2 = 1. \qquad (7)$$

The state of belief that some person $\theta$, who was not present in the list $\{\theta_1, \theta_2, \ldots \theta_m\}$, is the criminal means that information about $\theta$ is not present in the superposition (7). Moreover, this state is orthogonal to this superposition, since its prior probability is zero. However, the structure of $\psi_0$ may

---

[2] By demanding that each person is able to assign prior probabilities to all possible outcomes (rather than make a judgement about one particular outcome) we have to assume that, by performing a "prior-measurement", one does not irrevocably modify the initial belief state $\psi_0$ or at least $\psi_0$ can be perfectly reproduced to be used for further mental measurements. This may not sound plausible for physics, where the state collapses as the result of the measurement, but in studying cognition this assumption is natural. In physics there is assumed (at least, theoretically) a preparation procedure generating an ensemble of systems in the same state. A mental analog of this physical assumption about state preparation is that the brain, while solving a concrete problem, is able (and meant to) to return to the same belief state, after a judgment. However, for some mental contexts this assumption may be very restrictive. In principle, it is possible to proceed without it. But for the traditional Bayesian approach it is really important to start with the assignment of the prior probabilities, since they are explicitly involved in the update rule (3).

The crucial feature of the quantum scheme is that it is about the updates of states, not probabilities. Once the state has been updated, we can obtain the posterior probabilities. Therefore, in principle we can proceed without the explicit assignment of the prior probabilities $\pi(\theta)$ given by (5), meaning that the prior measurement of the $\Theta$-observable can be eliminated from the quantum scheme of probability updating. So, we can start simply with preparation of the initial belief state $\psi_0$ and its update resulting from gaining information with the aid of the *X*-observable, see below.

change dramatically after the *x*-evidence.

The new state (after the *x*-evidence) can have a component not belonging to *{θ₁, θ₂, ... θₘ}*. In this case the commissar would assign a nonzero probability (which may be quite substantial) that the person *θ*, which is not among *{θ₁, θ₂, ... θₘ}*, is the criminal.

One can imagine that together with obtaining the *x*-evidence, the observer (e.g., the commissar) is changing his own state, in an essentially non-classical manner. If a new option is born from influence of some new circumstances, the observer may find his horizons essentially widened by these circumstances, too.

Suppose, the commissar divides his belief among two suspects, so that his quantum belief state is $c_1 e_1 + c_2 e_2$, where vectors $e_1$ and $e_2$ are orthogonal (meaning that the options are mutually exclusive, only one or the other is true). In the present belief state, the options are also exhaustive, meaning that probabilities sum up to one, $|c_1|^2 + |c_2|^2 = 1$. Meanwhile, the actual state space may be three-dimensional, with basis vectors $e_1$, $e_2$, and $e_3$ (though the commissar is probably not aware of this), but the prior for $e_3$ is zero. Now, suppose we measure *x*-evidence, given by projector on the vector $(e_2 + e_3)/\sqrt{2}$. Probabilities about *x* are collected. But also, as a result of this measurement, the state of the commissar is changed, namely, it is projected on this vector. The component of the basis vector $e_3$ is no longer zero, and this is how the third option comes into consideration.

## 3 Experiment Investigation
### 3.1 Setup

We conducted two experiments to test whether participants' everyday/ intuitive decision making is consistent with Bayes's law for probability updating. In both cases, we created simple scenarios, based on a crime mystery. The experiments were matched in much of their detail, so we describe some of the common elements here. All participants were provided with a main story (approximately 620 words), which described a couple, John and Jane, their children (Chad and Cheryl) and a number of other persons, such as friends, John and Jane's gardener etc. The crucial bit of information was that Jane is fond of her jewelry, which, though valuable, is kept in her bedroom in an unsecured jewelry box. Participants are told that, on a particular Sunday evening, Jane discovers her jewelry is stolen. They are then asked to rate the probability that each of the persons introduced in the main story is the culprit. The persons were described so that some of them were slightly more likely to be guilty than others. However, in all cases, Chad (Jane's son) and John (Jane's husband) were initially completely beyond suspicion (this was directly verified in our data). That is, the main story was constructed so that the priors for Chad and John to be guilty were extremely low or zero.

After participants rated the suspects, they were given further information, which (unsurprisingly for the reader of this paper) provided a strong motive for stealing the jewels for Chad (in Experiment 1) and John (in Experiment 2). Participants were asked to rate the possible suspects again.

With this setup, we therefore have a situation such that a hypothesis with an extremely low prior suddenly acquires a high posterior. However, empirically, it is impossible to establish whether a prior is exactly zero vs. just extremely low. Therefore, our test of Cromwell's rule has to correspond to a test of whether probability updating is consistent with Bayes's law, $p(\theta|x) = \frac{p(x|\theta) \cdot p(\theta)}{p(x)}$ or $\frac{p(\theta|x)}{p(\theta)} = \frac{p(x|\theta)}{p(x)}$; put differently, according to Bayes's law, the ratio of the likelihoods is constrained by the ratio of the priors. To empirically test Bayes's law, we needed two kinds of information, (1) information on whether a suspect is guilty or not *a priori* and given the motive and (2) information on the motive *a priori* and given that the jewelry was stolen. Theoretically, it would have been ideal to collect data on all these probabilities within participants. However, asking participants to estimate the probabilities of conditionals and their corresponding reciprocals would have potentially been very confusing, so we adopted a between participants design, regarding the estimation of conditionals in the two directions.

### 3.2 Experiment 1 – Chad

#### 3.2.1 Participants

We tested 57 participants, all students at the University of California Irvine, who received fixed course credit for their time.

#### 3.2.2 Materials and procedure

The experiment was designed in Qualtrics and administered online. The experiment lasted for approximately 20 minutes. All participants received the main story featuring a family of John, Jane and their two children Chad and Cheryl, their two neighbors Matt and Mary, a cleaner, a gardener, and a burglar. Following this, there were 10 multiple choice questions (such as whether John has 1, 2, 3, or 4 children), which were meant to examine participants' basic knowledge of the story. The participants could not continue with the test until they answered all 10 questions right. Regarding safeguards that participants seriously engaged with the task, additionally, participants then received 6 catch questions testing probability intuitions as well as the understanding of the main story. Note, both the main story and all subsequent text that participants were presented with were accessible to participants throughout the experiment (that is, participants were told that this was not a memory test and they could take notes or make a copy of any screen). Four of the second group of 6 questions can be called 'easy catch' questions, and they corresponded to rather a trivial understanding of probability, such as what is the probability John has a son, when this fact was explicitly stated in the story. Two of these questions can be called 'hard catch' questions, and tested participants basic probability intuitions, for example, participants were asked "What is the probability that Mary has two daughters?" (a hard question) about the neighbor Mary who is known to be a mother of two, but the sex of her children was not explicitly provided in the main story.

After the catch questions, participants were told that Jane discovered her jewelry was missing, called the police, etc. Participants were given some facts that made it slightly likely that the cleaner or the gardener could have stolen the jewelry. Note also that one of the persons introduced in the main story was a local burglar, designed to be a fairly natural suspect. Following this information, participants were told to consider how likely all of the persons in the main story were to have stolen the jewelry – there were in total nine persons and participants were also allowed to enter a probability for 'other persons'. Note, participants were told to indicate their probability estimates for each suspect on a 0 to 100 scale and that their probabilities should sum up to 100 (this was checked automatically by the experimental software).

Following these ratings, the study went differently for two groups of participants. The first group was provided with a strong motive that Chad stole the jewelry. In brief, they were told that Chad wrecked a friend's car and needed money to repair it. They were then asked to rate the probability that each person had stolen the jewelry, as before. This was the last request in the experiment. This condition provided information for evaluating $\frac{Prob(Je|I)}{Prob(Je)}$, where *Je* denotes the event Chad stole the jewellery and *I* denotes the information that Chad wrecked the car and needed money to repair it.

In the second condition, after the first rating of the people, participants were given information about what was labeled a 'claim about Chad'. The claim was, basically, that Chad wrecked a friend's car. Participants were asked to rate the probability of this claim about Chad. In a subsequent screen, participants were informed that it is known for a fact that the jewelry was stolen by Chad and asked to update the probability of the claim about Chad. Thus, the second condition provides information that allows us to compute $\frac{Prob(I|Je)}{Prob(I)}$. At this point, it should be clear why we opted to adopt a between participants, rather than a within participants design: for participants who had been asked to estimate e.g., $Prob(Je|I)$, it would have been very confusing to subsequently estimate $Prob(I|Je)$ as well. Finally, note that all main parts of the text employed in the two conditions are shown in Appendix 1.

### 3.2.4 Results

In Table 1, we show the probability estimates regarding all people in the story, prior estimates (collected in both conditions of the experiment) as well as updated ones (collected in the first condition), following the information about Chad. The prior probability that Chad is guilty, *P(Je)*, was on average 1.5 % the updated probability *P(Je/I)* was on average 34 %, the prior probability of the claim about Chad, *P(I)*, was on average 40 % and the conditional probability of the claim about Chad, given that he had stolen the jewelry, *P(I/Je)*, was on average 79 %.

**Table 1. Results of Experiment 1 (Chad). Prior and updated (marked in grey) responses for each suspect. Shown are frequencies and mean probabilities. Standard deviations are given in parentheses.**

| Suspect | Frequencies, % | | | | | | | | | | | Averaged probability, % |
|---|---|---|---|---|---|---|---|---|---|---|---|---|
| | 0 | 0.5-10 | 11-20 | 21-30 | 31-40 | 41-50 | 51-60 | 61-70 | 71-80 | 81-90 | 91-100 | |
| John (prior) | 87 | 13 | 0 | 0 | 0 | 0 | 0 | 0 | 0 | 0 | 0 | 0.9 (2.7) |
| John (updated) | 87 | 13 | 0 | 0 | 0 | 0 | 0 | 0 | 0 | 0 | 0 | 0.9 (2.7) |
| Jane (prior) | 89 | 11 | 0 | 0 | 0 | 0 | 0 | 0 | 0 | 0 | 0 | 0.8 (2.5) |
| Jane (updated) | 87 | 13 | 0 | 0 | 0 | 0 | 0 | 0 | 0 | 0 | 0 | 1.0 (2.7) |
| Chad (prior) | 76 | 24 | 0 | 0 | 0 | 0 | 0 | 0 | 0 | 0 | 0 | 1.5 (3.2) |
| Chad (updated) | 0 | 27 | 7 | 30 | 7 | 13 | 3 | 3 | 7 | 0 | 3 | 34 (24) |
| Cheryl (prior) | 78 | 20 | 2 | 0 | 0 | 0 | 0 | 0 | 0 | 0 | 0 | 1.5 (3.5) |
| Cheryl (updated) | 80 | 20 | 0 | 0 | 0 | 0 | 0 | 0 | 0 | 0 | 0 | 1.1 (2.7) |
| Matt (prior) | 31 | 40 | 16 | 9 | 2 | 0 | 0 | 0 | 2 | 0 | 0 | 11 (13) |
| Matt (updated) | 34 | 50 | 13 | 0 | 0 | 0 | 0 | 3 | 0 | 0 | 0 | 7.8 (12) |
| Mary (prior) | 51 | 36 | 7 | 4 | 2 | 0 | 0 | 0 | 0 | 0 | 0 | 5.7 (7.8) |
| Mary (updated) | 50 | 40 | 7 | 0 | 3 | 0 | 0 | 0 | 0 | 0 | 0 | 5.1 (7.7) |
| Cleaner (prior) | 4 | 22 | 14 | 27 | 5 | 20 | 4 | 0 | 4 | 0 | 0 | 28 (19) |
| Cleaner (updated) | 10 | 33 | 27 | 23 | 7 | 0 | 0 | 0 | 0 | 0 | 0 | 16 (11) |
| Gardener (prior) | 4 | 24 | 22 | 38 | 5 | 5 | 2 | 0 | 0 | 0 | 0 | 22 (13) |
| Gardener (updated) | 10 | 44 | 23 | 20 | 3 | 0 | 0 | 0 | 0 | 0 | 0 | 14 (9.5) |
| Burglar (prior) | 15 | 24 | 18 | 18 | 5 | 11 | 7 | 0 | 2 | 0 | 0 | 24 (20) |
| Burglar (updated) | 20 | 30 | 27 | 13 | 4 | 0 | 3 | 0 | 3 | 0 | 0 | 17 (17) |
| Others (prior) | 67 | 22 | 4 | 4 | 0 | 3 | 0 | 0 | 0 | 0 | 0 | 5.1 (11) |
| Others (updated) | 76 | 17 | 0 | 0 | 7 | 0 | 0 | 0 | 0 | 0 | 0 | 3.4 (10) |

We seek to test against the null hypothesis that $\frac{Prob(Je|I)}{Prob(Je)} = \frac{Prob(I|Je)}{Prob(I)}$, as required by Bayes's law of probability updating. Intuitively, the reason why Bayes's law may be violated in this case is that we were expecting a huge increase in the probability that Chad is guilty, from the prior estimate *P(Je)* to the estimate *P(Je/I)*, following the information that Chad had wrecked the car. Note first that given expectations for low (or zero) priors, we computed for each participant either $\frac{Prob(Je)}{Prob(Je|I)}$ or $\frac{Prob(I)}{Prob(I|Je)}$, depending on condition. Still, one participant produced $Prob(Je|I) = 0$ (out of 31) and another $Prob(I|Je) = 0$ (out of 26) and these were eliminated from further consideration (their results cannot be used to evaluate Bayes's law). We used the MATLAB software package to calculate *z* and *p* values.

A non-parametric Mann-Whitney U-test of $\frac{Prob(Je|I)}{Prob(Je)}$ vs. $\frac{Prob(I|Je)}{Prob(I)}$ was found highly significant, $z=4.6$, $p=5 \cdot 10^{-6}$ (U=120, $n_1$=30, $n_2$=25), allowing us to reject the null hypothesis that these two quantities are equal. Next, recall, the easy catch questions provided a measure of participants' attention paid to the story plus some trivial understanding of the probability concept (max possible score 4) and the hard catch questions a measure of participants' probabilistic intuitions (max possible score 2). We ran the U-test comparison once more, but excluding all participants with an easy catch score of 2 or less (23 participants were excluded like this); the test was significant again, $z=-2.8$, $p=.006$, (U=59, $n_1$=16, $n_2$=16). Overall, across a number of checks, there was strong evidence that probability updating was not consistent with Bayes's law.

### 3.3 Experiment 2 – John

#### 3.3.1 Participants
We tested 58 participants, all students at the University of California Irvine, who received fixed course credit for their time.

#### 3.3.2 Materials and procedure
Most aspects of this experiment are the same as for Experiment 1 – Experiment 2 was meant to be a simple replication of Experiment 1. Therefore, we only describe the ways in which Experiment 2 differed from Experiment 1.

In the first condition, after participants submitted the prior estimates of the probability that each person stole the jewelry, participants were provided with a strong motive for John. They were told that John had a severe, secret gambling problem and he owed a lot of money. They were also told that if Jane were to find out about John's debts, she would likely divorce him. So, this motive was meant to make John a very likely suspect. The first condition ended as in Experiment 1, with participants being asked to update the probability that each suspect was guilty, following this new information about John.

In the second condition, following prior evaluation, participants were told about the above 'claim for John' and were asked to evaluate its probability, *P(I)*. They were then told that the jewelry was definitely stolen by John and they were asked to rate again the probability of the claim about John, given John had stolen the jewelry *P(I|Je)*. In Appendix 2, the interested reader can see the text for Experiment 2 that is different, compared to Experiment 1.

#### 3.3.2 Results
We proceed as for Experiment 1. Table 2 shows the estimates for the prior and updated probability for each person of the story stealing the jewels. The prior probability that John has stolen the jewels, *P(Je)* was on average 1.6 %, the updated *P(Je|I)* was on average 39 %, the prior probability for the claim about John, *P(I)* was on average 41 %, and the conditional probability for the claim, given that John was known to have stolen the jewelry, was on average 83 %.

**Table 2. Results of Experiment 2 (John). Prior and updated (marked in grey) responses for each suspect. Shown are frequencies and mean probabilities. Standard deviations are given in parentheses.**

| Suspect | Frequencies, % | | | | | | | | | | | Averaged probability, % |
|---|---|---|---|---|---|---|---|---|---|---|---|---|
| | 0 | 0.5-10 | 11-20 | 21-30 | 31-40 | 41-50 | 51-60 | 61-70 | 71-80 | 81-90 | 91-100 | |
| John (prior) | 78 | 20 | 0 | 2 | 0 | 0 | 0 | 0 | 0 | 0 | 0 | 1.6 (4.1) |
| John (updated) | 0 | 11 | 25 | 18 | 14 | 7 | 7 | 4 | 3 | 0 | 11 | 39 (28) |
| Jane (prior) | 85 | 11 | 2 | 0 | 0 | 0 | 0 | 0 | 0 | 0 | 2 | 2.9 (13.8) |
| Jane (updated) | 89 | 11 | 0 | 0 | 0 | 0 | 0 | 0 | 0 | 0 | 0 | 0.8 (2.6) |
| Chad (prior) | 78 | 20 | 0 | 2 | 0 | 0 | 0 | 0 | 0 | 0 | 0 | 1.6 (4.7) |
| Chad (updated) | 82 | 18 | 0 | 0 | 0 | 0 | 0 | 0 | 0 | 0 | 0 | 1.3 (3.2) |
| Cheryl (prior) | 76 | 24 | 0 | 0 | 0 | 0 | 0 | 0 | 0 | 0 | 0 | 1.3 (2.7) |
| Cheryl (updated) | 86 | 14 | 0 | 0 | 0 | 0 | 0 | 0 | 0 | 0 | 0 | 1.2 (3.2) |
| Matt (prior) | 38 | 38 | 18 | 6 | 0 | 0 | 0 | 0 | 0 | 0 | 0 | 7.5 (8) |
| Matt (updated) | 46 | 47 | 7 | 0 | 0 | 0 | 0 | 0 | 0 | 0 | 0 | 4.6 (5.3) |
| Mary (prior) | 40 | 36 | 20 | 2 | 0 | 2 | 0 | 0 | 0 | 0 | 0 | 7.4 (9) |
| Mary (updated) | 50 | 39 | 11 | 0 | 0 | 0 | 0 | 0 | 0 | 0 | 0 | 4.4 (5.6) |
| Cleaner (prior) | 15 | 16 | 25 | 31 | 7 | 4 | 0 | 0 | 2 | 0 | 0 | 20 (14) |
| Cleaner (updated) | 25 | 32 | 14 | 25 | 0 | 0 | 4 | 0 | 0 | 0 | 0 | 14 (13) |
| Gardener (prior) | 11 | 22 | 24 | 34 | 7 | 2 | 0 | 0 | 0 | 0 | 0 | 18 (11) |
| Gardener (updated) | 22 | 43 | 14 | 21 | 0 | 0 | 0 | 0 | 0 | 0 | 0 | 12 (9) |
| Burglar (prior) | 7 | 15 | 14 | 15 | 7 | 27 | 2 | 2 | 7 | 0 | 4 | 36 (25) |
| Burglar (updated) | 18 | 39 | 18 | 7 | 7 | 4 | 0 | 4 | 0 | 3 | 0 | 19 (21) |
| Others (prior) | 73 | 15 | 5 | 7 | 0 | 0 | 0 | 0 | 0 | 0 | 0 | 4.0 (8.3) |
| Others (updated) | 75 | 18 | 0 | 7 | 0 | 0 | 0 | 0 | 0 | 0 | 0 | 3.2 (7.5) |

Using the notation of Experiment 1, we excluded participants for whom $\frac{Prob(Je)}{Prob(Je|I)}$ or $\frac{Prob(I)}{Prob(I|Je)}$ were uncomputable. This was the case for two participants in the first condition (out of 30) and one in the second (out of 28). The two quantities on either side of Bayes's law were significantly different, $z=4.7$, $p=2\cdot10^{-6}$ (U=648, $n_1$=28, $n_2$=27), thus we could not sustain the null hypothesis that participants' probability updating was consistent with Bayes's law. The same conclusion was reached when we ran the same U-test, but excluding all participants with an easy catch score of 2 or less (23 participants excluded; $z=2.9$, $p=.004$, U=202, $n_1$=17, $n_2$=15).

## 3. QPT application

In this section we illustrate how QPT can accommodate large jumps in probability updating. The human behavior we are interested in corresponds *just* to the process of probability updating. Thus, there is no detailed model to be constructed, instead we specify the representation and show that QPT can reproduce the observed probability updating. This is not possible with CPT, as shown in the results sections of the two experiments.

We illustrate QPT probability updating in two ways, a coarse grained one and a fine grained one. Starting with the former, let us number the basis vectors as follows: $e_1=|\Theta=Chad\rangle$, $e_2=|\Theta=the\ burglar\rangle$, $e_3=|\Theta=the\ cleaner\rangle$, $e_4=|\Theta=the\ gardener\rangle$, $e_5=|\Theta=others\rangle$. In what follows, we confine our consideration to this five-dimensional case. Conveniently, in the above basis, the probabilities for the different hypotheses are given as the squared norm of coefficients before the basis vectors. Let us take the prior state in the simplest form, such that the probabilities are those given in Table 3:

$$\psi_0 = \sqrt{0.5}e_2 + \sqrt{0.2}e_3 + \sqrt{0.2}e_4 + \sqrt{0.1}e_5. \tag{8}$$

**Table 3. Prior beliefs for each hypothesis**

| $\Theta$=John | $\Theta$=Jane | $\Theta$=Chad | $\Theta$=Cheryl | $\Theta$=Matt | $\Theta$=Mary | $\Theta$=the gardener | $\Theta$=the cleaner | $\Theta$=the burglar | $\Theta$= others |
|---|---|---|---|---|---|---|---|---|---|
| 0 % | 0 % | 0 % | 0 % | 0 % | 0 % | 20 % | 20 % | 50 % | 10 % |

Note that $e_1$ does not appear in the above equation because the probability amplitude for the Chad vector is zero. The other amplitudes just correspond to the observed data. Given this state vector, $\psi_0$, we can reproduce all the prior probabilities, by projecting $\psi_0$ on the mutually orthogonal basis vectors defined above, such as $|\Theta=Chad>$, $|\Theta=the\ burglar>$,... The corresponding projectors $E_\theta$, (one projector for each hypothesis; Equation 5) are given by $E_\theta=|e_i><e_i|$ and lead to the corresponding probabilities in Table 3 through $<E_\theta\psi_0,\psi_0>$. Note, $<vector|$ is the complex conjugate of $|vector>$. Finally, state $\psi_0$ represents the cognitive state of a typical or average participant in the experiment, after the first step of evaluating suspect probabilities, but we are not concerned with the cognitive state before this step. This state is meant to reproduce the response statistics across the sample, on average. As our objective is focused on probability updating behavior observed in the sample as a whole, we did not pursue modelling of individual differences (with future extensions, this could be done with the formalism of density matrices).

What is the effect of new information about Chad on the mental state $\psi_0$? Mostly, it advances the Chad hypothesis without affecting the other hypotheses. We know the specific effect this new information has on the mental state, since we know how the prior probabilities change, after the new information. Thus, if the initial state is $\psi_0$, as above, then we know that the effect of the new information would be to project $\psi_0$ to a new state $\psi_I$ with coefficients such that their squared norm gives the updated probabilities. This new state $\psi_I$ simply corresponds to the observed updated probabilities, after participants received the new information; basically, the probability for Chad has "jumped" from zero to a substantial number, while all other probabilities decreased. We can identify a projector operator that projects $\psi_0$ to $\psi_I$ (note that $\psi_I$ and a projector are not unique; however, all projectors have to be symmetric and equal to their square) and an example is:

$$F_I = \begin{pmatrix} 23/32 & -9/32 & -3\sqrt{1.5}/16 & -3\sqrt{2}/16 & 0 \\ -9/32 & 23/32 & -3\sqrt{1.5}/16 & -3\sqrt{2}/16 & 0 \\ -3\sqrt{1.5}/16 & -3\sqrt{1.5}/16 & 13/16 & -\sqrt{3}/8 & 0 \\ -3\sqrt{2}/16 & -3\sqrt{2}/16 & -\sqrt{3}/8 & 3/4 & 0 \\ 0 & 0 & 0 & 0 & 0 \end{pmatrix}. \quad (9)$$

Using this projector, our initial state (7) is updated to

$$\psi_I = \sqrt{0.65}e_1 + \sqrt{0.3}e_2 + \sqrt{0.04}e_3 + \sqrt{0.01}e_4. \quad (10)$$

Finally, to obtain the probability of each person being the thief, we again apply the projector $E_\theta$ for a particular suspect $\theta$ as shown in Eq (5). The probability for Chad has "jumped" from zero to 65%, while all other probabilities decreased. Note, a proper cognitive model of this decision making situation would focus more on how the different kinds of evidence impact on beliefs about states. However, as noted, here we only want to illustrate how QPT can accommodate large increases in probability; Equation (9) effectively constitutes an existence proof that QPT can. Note, for simplicity, we took the last row of our projector to be zero. Naturally, this resulted in zero updated probability of $e_5$.

In the experiment we observed the following: when updating from $P(Je)$ to $P(Je|I)$ (both in Experiments 1 and 2), 38 % of the participants changed at least one of their estimates from non-zero to zero; at the same time, 44 % of the participants left at least one of their estimates unchanged.

If one seeks a more fine-grained description of the results, then we can easily extend the above approach. Let us consider a more fine-grained basis, corresponding to the suspects in the order they appear in Tables 1, 2, and 3: $e_1=|\Theta=John>$, $e_2=|\Theta=Jane>$, $e_3=|\Theta=Chad>$, $e_4=|\Theta=Cheryl>$, $e_5=|\Theta=Matt>$, $e_6=|\Theta=Mary>$, $e_7=|\Theta=the\ gardener>$, $e_8=|\Theta=the\ cleaner>$, $e_9=|\Theta=the\ burglar>$, $e_{10}=|\Theta=others>$. For Experiment 1, the initial state can be chosen as:

$$\psi_0 = \frac{1}{10}\left(\sqrt{0.9}e_1 + \sqrt{0.8}e_2 + \sqrt{1.5}e_3 + \sqrt{1.5}e_4 + \sqrt{11}e_5 + \sqrt{5.7}e_6 + \sqrt{28}e_7 + \sqrt{22}e_8 + \sqrt{24}e_9 + \sqrt{5.1}e_{10}\right). \quad (11)$$

It can be readily confirmed that this state vector reproduces all the estimates for the prior probabilities for all suspects, as in Table 1 (the unshaded rows). The impact of the new information on $\psi_0$ can be given through the projector $F_I = |f\rangle\langle f|$, where

$$f = \frac{1}{10}\left(1.012e_1 + 1.066e_2 + 6.217e_3 + 1.118e_4 + 2.978e_5 + 2.408e_6 + 4.265e_7 + 3.99e_8 + 4.396e_9 + 1.966e_{10}\right)$$

(12)

For illustration, we compute the projector explicitly below:

$$F_I = \frac{1}{100}\begin{pmatrix} 1.024 & 1.079 & 6.292 & 1.131 & 3.014 & 2.437 & 4.316 & 4.038 & 4.449 & 1.990 \\ 1.079 & 1.137 & 6.627 & 1.192 & 3.175 & 2.567 & 4.546 & 4.253 & 4.686 & 2.096 \\ 6.292 & 6.627 & 38.651 & 6.951 & 18.514 & 14.971 & 26.516 & 24.806 & 27.330 & 12.223 \\ 1.131 & 1.192 & 6.951 & 1.250 & 3.329 & 2.692 & 4.768 & 4.461 & 4.915 & 2.198 \\ 3.014 & 3.175 & 18.514 & 3.329 & 8.868 & 7.171 & 12.701 & 11.882 & 13.091 & 5.855 \\ 2.437 & 2.567 & 14.971 & 2.692 & 7.171 & 5.798 & 10.270 & 9.608 & 10.586 & 4.734 \\ 4.316 & 4.546 & 26.516 & 4.768 & 12.701 & 10.270 & 18.190 & 17.017 & 18.749 & 8.385 \\ 4.038 & 4.253 & 24.806 & 4.461 & 11.882 & 9.608 & 17.017 & 15.920 & 17.540 & 7.844 \\ 4.449 & 4.686 & 27.330 & 4.915 & 13.091 & 10.586 & 18.749 & 17.540 & 19.325 & 8.643 \\ 1.990 & 2.096 & 12.223 & 2.198 & 5.855 & 4.734 & 8.385 & 7.844 & 8.643 & 3.865 \end{pmatrix} \quad (13)$$

The resulting updated state $\psi_1 = \langle f|\psi_0\rangle |f\rangle$ is

$$\psi_1 = \frac{1}{10}\left(\sqrt{0.9}e_1 + e_2 + \sqrt{34}e_3 + \sqrt{1.1}e_4 + \sqrt{7.8}e_5 + \sqrt{5.1}e_6 + \sqrt{16}e_7 + \sqrt{14}e_8 + \sqrt{17}e_9 + \sqrt{3.4}e_{10}\right) \quad (14)$$

As previously, it is clear that projecting $\psi_1$ to a basis vector $e_i$ will yield the corresponding posterior probability, after participants had processed the new information regarding the theft case (as specified in $i$th shaded row of Table 1).

Similarly, for Experiment 2, the initial state can be set to:

$$\psi_0 = \frac{1}{10}\left(\sqrt{1.6}e_1 + \sqrt{2.9}e_2 + \sqrt{1.6}e_3 + \sqrt{1.3}e_4 + \sqrt{7.5}e_5 + \sqrt{7.4}e_6 + \sqrt{20}e_7 + \sqrt{18}e_8 + \sqrt{36}e_9 + \sqrt{4}e_{10}\right), \quad (15)$$

Thus, prior probabilities coincide with those in the unshaded rows of Table 2. The projector corresponding to receiving the new information can be taken to be of the form $|f\rangle\langle f|$, where

$$f = \frac{1}{10}\left(6.779e_1 + 0.971e_2 + 1.238e_3 + 1.189e_4 + 2.328e_5 + 2.277e_6 + 4.062e_7 + 3.761e_8 + 4.732e_9 + 1.942e_{10}\right)$$

(16)

The resulting state $\psi_1 = \langle f|\psi_0\rangle |f\rangle$ is then

$$\psi_1 = \frac{1}{10}\left(\sqrt{39}e_1 + \sqrt{0.8}e_2 + \sqrt{1.3}e_3 + \sqrt{1.2}e_4 + \sqrt{4.6}e_5 + \sqrt{4.4}e_6 + \sqrt{14}e_7 + \sqrt{12}e_8 + \sqrt{19}e_9 + \sqrt{3.2}e_{10}\right). \quad (17)$$

Projecting $\psi_1$ to a basis vector $e_i$ will yield the corresponding probability equal to the one specified in $i$th shaded row of Table 1.

Finally, we can consider what are the implications regarding psychological process, from the application of QPT in this example. We can be guided by Dubois and Lambert-Mogiliansky (2015). Their argument appeals to the incompatibility of perspectives in the mind and the process of learning. For example, from the perspective of the initial information about possible suspects, John appears as an extremely unlikely suspect. In technical terms, the basis corresponding to the initial perspective makes it very unlikely that John is a suspect. When the new information about John's gambling becomes available (Experiment 2), the mind shifts perspectives regarding John and the possibility that he is guilty. This new perspective incorporates information about how he is perhaps not a bad man, but his weakness in gambling now endangers his marriage. With this new perspective, the basis set for evaluating the possible guilt of different suspects likewise changes and the projection onto the subspace corresponding to John's guilt becomes large. Importantly, the new perspective produces probabilities that are not linearly related to the priors and so QPT can accommodate changes from priors to posteriors not possible classically, as well as interference effects etc.

## 4. Discussion

Experiments 1 and 2 are very similar and demonstrated very similar results. About 20% of the participants updated their zero or nearly zero priors to high confidence (more than 50%) upon receiving new information about John or Chad, demonstrating a strong deviation from Cromwell's rule. Probably, the zeros initially entered by participants for some suspects are but a shorthand for some small number, like 0.00001. Participants were allowed but not encouraged to input long decimals. Still, the results cannot be accommodated by Bayesian updating. Moreover, an assumption that people can have such a refined scale of probability may not be very plausible.

An attempt to reconcile these results with classical Bayesian updating may lie in direct observation (experimental collection of the relevant data) of all ten conditional probabilities. Denote the ten possible options $\Theta=\theta_1$, $\Theta=\theta_2$, ..., in general, $\Theta=\theta_i$, where by $\Theta$ we mean "the thief", and by $\theta_i$ all possible suspects, for example, $p(\Theta=\theta_1)$ is the prior probability that John is the thief. In principle,

$$p(I) = \sum_{i=1}^{10} p(I | \Theta = \theta_i) p(\Theta = \theta_i). \qquad (18)$$

In the present work, participants were asked to directly evaluate the probability $p(I)$. Alternatively, we could have asked them about all $p(I|\Theta=\theta_i)$ on the right hand side of (18) and then calculate $p(I)$. Would this lead to smaller violation of the Bayesian updating? If yes, this would mean that the present experiments can be interpreted as an unpacking effect (Tversky & Kohler, 1994), a kind of violation of the law of total probability, in the form (18), or a conjunction effect, where $p(I)$ is harder to estimate than $p(I|\Theta=\theta_i)$ and, hence, estimates for $p(I)$ are smaller than $p(I|\Theta=\theta_i)$. For the moment, this is only a hypothesis, which shall be put to experimental test in the future.

The possibilities $p(\Theta=\theta_i)$ are always assumed to be mutually exclusive and exhaustive, meaning that exactly one suspect is actually the thief. Probably, these are the features that make Bayesian updating problematic in decision making. As noted in the Introduction, in earlier experiments violating Bayesian updating, Shaferian representation of beliefs was quite successfully applied (staying in the framework of CP theory). This theory aims to relax assumptions of exhaustiveness and mutual exclusiveness of options (and account for experimental violation of these assumptions), by working with groups of options. For example, one option may enter a few groups, so the groups cannot be thought of as mutually exclusive.

Shafer's idea is that upon the new information in favor or against one option, e.g. $\theta_1$, probability is reallocated within each group containing $\theta_1$, so that structure of the groups may become finer, which finally leads to a decision. Interestingly, estimates of the other options, not initially grouped with $x_1$, remain the same. Could our results be accounted for by a generalized version of CP theory, such as the one proposed by Shafer?

A common prior probability distribution we observed in our experiment looks like Table 3: the gardener 20%, the cleaner 20%, the burglar 50%, others 10%, other six personages (four of the family plus two neighbors) 0% each.

Suppose that the actual belief system was a grouping (for simplicity, the subsets are mutually exclusive) as shown in Table 4.[3] One may interpret this grouping as a reluctance to assign a zero prior to any hypothesis. Also, this saves computational capacity. A participant may think that the most obvious suspect is the burglar and that there are two other, equally probable, suspects; then, one can group whoever is left into the least probable belief set.

**Table 4. Prior belief representation.**

| {the family, the neighbors, others} | {the gardener, the cleaner} | {the burglar} |
|---|---|---|
| 10 % | 40 % | 50 % |

Now, the evidence pointing towards Chad or John as the possible criminals does not entail Bayesian updating on a zero prior, rather the new key suspect can reclaim all 10% of belief belonging to his group. This would nicely correspond to this particular result of our experiment (updating from zero to a significant probability). Still, other estimates, which remain unchanged in this (rather simple) Shaferian model, were found to significantly change in the experiment. According to Shafer's theory, the (classical) belief state updated on the incriminating information about Chad, would be as shown in Table 5.

**Table 5. Updated belief representation.**

| {Chad} | {the family excluding Chad, the neighbors, others} | {the gardener, the cleaner} | {the burglar} |
|---|---|---|---|
| 10 % | 0 % | 40 % | 50 % |

This does not coincide with the results of our experiment (nor with our intuition) and indicates that QPT approach is the more effective one for our problem. Of course, a more sophisticated Shaferian representation can be constructed to better accommodate the results, but we shall not address this in detail. Let us only note that two opposite directions – adding Chad to more and more belief groups (to all of them, in the extreme case) or separating him from other suspects (in the extreme case, to the group of his own) finally result in a standard Bayesian updating scheme.

We have shown that Shafer theory does not describe our data (at least, in a straightforward manner), in contrast to QPT, as our illustration, presented in Section 4, shows. One feature that makes QPT promising is that quantum belief states are much richer (contain much more information) than classical belief states.

To summarize, we suggest a method from QPT to describe our experimental data, which, we found, violates classical Bayesian updating. The latter is based on probability of a joint event
Prob(hypothesis|data)·Prob(data)=Prob(data|hypothesis)·Prob(hypothesis)=Prob(hypothesis&data).(19)
Finding a direct analogue to this expression consistent with QPT is hardly feasible. To mention one problem, most often a combination (a product) of two projectors is not a projector. Still, some attempts can be found in the literature on quantum physics (Steinberg, 1995), but the limitations must always be in sight. We can safely calculate Prob(data) as $\|\langle\text{Projector data} | \text{belief state}\rangle\|^2$. For example, in our example from Section 4, the hypothesis under consideration is $\Theta=\theta_1$. Applying operator (9) to prior state $\psi_0$ in form (8), and taking the norm of the resulting vector

$$F_I\psi_0 = -0.42e_1 + 0.29e_2 + 0.1e_3 + 0.05e_4 , \qquad (20)$$

we get Prob(data)=0.27 (27%). Now, guided by formula (5) we apply the projector $E_1=|e_1\rangle\langle e_1|$ to $\psi_I = F_I \psi_0/0.27^{1/2}$ ($\psi_I$ is normed $F_I \psi_0$) and we get Prob (hypothesis|data)=0.42/0.27=0.65 (65%). Still, Prob(data|hypothesis)=0 as the result of applying operator (9) to zero vector, which, in turn, is the result of projecting $\psi_0$ to $e_1$. Using these values in formula (19), we get 0.65·0.27=0·0, that is, a violation of Bayesian updating again, which is due to noncommutativity of operators (9) and $E_1$.

---
[3] Even if beliefs are assigned to groups of options, presumably probability of a single option may be extracted at request.

Noncommutativity often leads to such order effects and, indeed, our data can be interpreted as an order effect

$$p(Je)\,p(I|Je) = p(Je \cap I) \neq p(I \cap Je) = p(I)\,p(Je|I) \quad (21)$$

5. **Conclusion**

We experimentally observed violations of classical, Bayesian updating of belief. As shown, updating on strong evidence can lead to a dramatic increase of confidence (from zero, practically denying the possibility) to almost complete confidence. We explain how and why quantum probability theory can be applied to describe the experimental results and resolve the zero-prior trap, in a way which is probably more efficient than following Cromwell's rule (applying only non-zero and non-one probabilities to all the options). The present work provides new insights potentially applicable in the experimental and theoretical study of the phenomenon of creativity, which can be interesting not only for cognitive psychology but also for more applied subjects, such as artificial intelligence.

**Acknowledgements**

IB was supported by a Marie Sklodowska-Curie Individual Fellowship, grant agreement 696331. EMP was supported by Leverhulme Trust grant RPG-2013-004 and by Air Force Office of Scientific Research (AFOSR), Air Force Material Command, USAF, grant FA 8655-13-1-3044. The U.S. Government is authorized to reproduce and distribute reprints for Governmental purpose notwithstanding any copyright notation thereon. JST was supported by NSF grant SES-1326275. IB and AK were supported by the grant Mathematical Modeling of Complex Hierarchic systems of Faculty of Technology, Linnaeus University.

**Appendix 1. The text employed in Experiment 1 - Chad.**

**Main story (the same for groups one and two):**

John and Jane are a fairly wealthy family, living in a large house. Their house is located on a quiet road, in a leafy suburb of a middle-sized town. They have been living there for several years and know their neighbors quite well. John and Jane have two young children, Chad and Cheryl, who still go to primary school. Many of their neighbors have children as well and one often sees a crowd of children playing on their road.

John is a medical doctor, specialized on pediatrics. He works at a local hospital. He seems to be good at his job and he enjoys a good reputation in the local community.

Jane works as an assistant in a cosmetics shop. The shop is a small business that has been doing quite well. Jane is very meticulous about her looks and she spends nearly all her money on jewelry. She has an impressive collection of jewelry, including several diamond rings and golden items. She keeps her jewelry, in a jewelry box, in the master bedroom of the house. She is very emotionally attached to her jewelry. The jewelry box has a lock, but she often just forgets the key in the lock.

Chad is 11 years old. He is often seen playing ball, with other children on his street. He is sociable. He likes school and does all right at school.

Cheryl is 9 years old. She likes drawing and takes an evening art class. She would like to become a doctor, like her dad. She is very close friends with some other children in the neighborhood.

Sibyl is a cleaner who comes to John and Jane's house once every week, usually on a Saturday morning, while everybody is out. She seems conscientious and has been working for John and Jane for many years. She is a single mother who lives just with her one child. She is 32 years old and she sometimes feels that her life is going nowhere. Recently, her child has been sick, but she has not been able to afford to take her to the hospital for proper care.

Richard is the gardener who comes to John and Jane's house once a week as well. The days on which he comes vary, since he avoids coming when it is rainy. Richard is hard working and has, so far, seemed content with this lot in life. He has access to the backdoor of the house. He has recently started going out with a girlfriend from a very wealthy family. He is very much in love with her and wants to impress her.

Matt and Mary are neighbors. They also have a large house and are good friends with John and Jane. They are not as wealthy as John and Jane and recently they had to trade their car for a smaller one. They also have two children.

Matt is a lawyer. He specialized on criminal law. However, he has been finding his work stressful and has not been taking many new cases. He would like to find, somehow, a chunk of money to repay the remaining of the mortgage on his house and then sit back and reevaluate his career.

Mary does not work. She had her two children when she was fairly young. Now, the children are older and need less attention, but she never managed to get a career going. She is now mostly happy to look after the house and pass her time with various hobbies.

Andrew is a burglar operating in the town where John and Jane live. He is very professional and likes to do his research before attempting to enter a house. He prefers to steal smaller valuables, since these are easier to sell.

**(At this point participants see the following test questions)**

Who is John's wife?
- o  Sibyl
- o  Katy
- o  Cleopatra

How many children does John have?
- o 1
- o 2
- o 3
- o 4

Where does Jane work?
- o At a cosmetics shop.
- o At a supermarket.
- o At a law firm.
- o At a nuclear station.

How old is Cheryl?
- o 7
- o 8
- o 9
- o 10

Who is Cheryl's brother?
- o Richard.
- o Chad.
- o John.
- o Matt.

What is the name of the cleaner?
- o Richard.
- o Mary.
- o Sibyl.
- o Jane.

When does the cleaner usually come?
- o Mondays.
- o Thursdays.
- o Saturdays.
- o We don't know.

What is the name of gardener?
- o Richard.
- o Sibyl.
- o Matt.
- o Mary.

What does Matt do?
- o He is a cleaner.

- He is a secret agent.
- He is a doctor.
- He is a lawyer.

How many children do Matt and Mary have?
- 1
- 2
- 3
- 4

**(At this point participants see the following catch questions)**

What is the probability that Jane has a daughter?
- 0%
- 25%
- 50%
- 75%
- 100%

What is the probability that Sybil has a daughter?
- 0%
- 25%
- 50%
- 75%
- 100%

What is the probability that Sybil has a son?
- 0%
- 25%
- 50%
- 75%
- 100%

What is the probability that Mary has two daughters?
- 0%
- 25%
- 50%
- 75%
- 100%

What is the probability that Mary has no daughters?
- 0%
- 25%
- 50%
- 75%
- 100%

What is the probability that Cheryl has a sister?
- 0%
- 25%
- 50%

- 75%
- 100%

**At this point participants see the information about theft**

**Theft**

It's Sunday evening, after dinner. Chad and Cheryl are in their rooms, reading, before going to sleep. John is having a shower. Jane is getting ready for bed. She opens her jewelry box to discover that all her jewelry is missing! John and Jane quickly call the police. These are the facts: The last time Jane checked her jewelry was Friday evening. Saturday morning was a sunny day. Sibyl (the cleaner) was due to come (she did). Richard (the gardener) came too, at some point in the morning, though he did not see Sibyl. All the family (John, Jane, Chad, Cheryl) was out Saturday morning. John went for a bicycle ride; Jane was working; Char and Cheryl visited some friends. They came back home at different times on Saturday, though no one came before 1:00 pm and they did not see either Sibyl or Richard at the house. In the evening of Saturday, John and Jane had invited Matt and Mary for dinner.

**At this point participants are asked to rate the suspects.**

**The above part of the scenario, including the prior rating, is the same for Experiment 1 (both groups) and Experiment 2 (both groups).**

**We continue with Experiment 1. The remainder of the test goes differently for the two groups.**

**The <u>first group</u> of participants is given new information as follows.**

Chad is in big trouble. One evening in the last few weeks, he and a friend of his "borrowed" the car of the friend's dad. They only wanted to drive around the block, but things got a bit out of control, and they ended up crashing the car. Fortunately, it was possible to have it repaired and returned promptly. But now Chad needs to settle the bill for the repairs or the mechanic threatens he will go to Chad's family or the police. Jane's jewelry would cover the cost.

**At this point participants of the <u>first group</u> are asked to rate the suspects again and this is the end of the test for them.**

**For the <u>second</u> between participants condition, we had main story, catch questions, information about the theft, prior probability estimates, and then the second group is presented with:**

<u>Claim about Chad:</u> The claim is that is that Chad is in big trouble. Allegedly, one evening in the last few weeks, he and a friend of his "borrowed" the car of the friend's dad. They only wanted to drive around the block, but things got a bit out of control, and they ended up crashing the car. Fortunately, it was possible to have it repaired and returned promptly. But now Chad needs to settle the bill for the repairs or the mechanic threatens he will go to Chad's family or the police. Jane's jewelry would cover the cost.

**At this point <u>the second group</u> participants are asked to rate the probability of the claim about Chad. Then, the second group is given new information:**

After some investigation, it turns out that (for some reason) Chad stole the jewelry. This is completely certain and Chad himself admitted that he stole the jewelry.

**At this point participants of the <u>second group</u> are reminded of the Claim about Chad, they are presented with the text of the claim and are asked to rate the probability of the claim once more. This is the end of the test for the second group.**

**Appendix 2. The text employed in Experiment 2 - John.**

The main story, catch question, and the text informing that the theft took place are the same. Then, the participants were asked to estimate prior probabilities that each of the personages is the thief.

**Then, remainder of the test goes differently for the two groups of participants in Experiment 2.**

**The <u>first group</u> of participants is given new information as follows**

It turns out that John has been having a severe gambling problem. He recently lost a very large amount of money. He is worried that if his wife were to find out about his debt, she will divorce him and he will lose everything. Jane's jewelry will cover all of John's gambling debts.

**At this point participants of the <u>first group</u> are asked to rate the suspects again and this is the end of the test for them.**

**For the <u>second</u> between participants condition, we had main story, catch questions, information about the theft, prior probability estimates, and then the second group is presented with:**

<u>**Claim about John:**</u> The <u>claim</u> is that John has been having a severe gambling problem and that he recently lost a very large amount of money. <u>Allegedly</u>, he is worried that if his wife were to find out about his debt, she will divorce him and he will lose everything. Jane's jewelry will cover all of John's gambling debts.

**At this point <u>the second group</u> participants are asked to rate the probability of the claim about John. Then, the second group is given new information:**

After some investigation, it turns out that (for some reason) John stole the jewelry. This is completely certain and John himself admitted that he stole the jewelry.

**At this point participants of the <u>second group</u> are reminded of the Claim about John, they are presented with the text of the claim and are asked to rate the probability of the claim once more. This is the end of the test for the second group.**